\documentclass[12pt]{article}
\usepackage{epsf}

\begin{document}
\begin{center}
{\Large \bf EXPERIMENTAL DATA ON THE SINGLE SPIN ASYMMETRY AND THEIR
INTERPRETATIONS BY THE CHROMO-MAGNETIC STRING MODEL}
\vspace*{1cm}
\textsc{  S.B. Nurushev$^{a}$ and M.G. Ryskin$^{b}$}

\vspace*{1cm} $^a$ Institute for High Energy Physics,
142281 Protvino, Moscow region, Russia\\
$^b$ Petersburg
Nuclear Physics Institute, 188300 Gatchina, St. Petersburg,
Russia
\end{center}

\vspace*{1cm}

\begin{abstract}

An attempt is made to interpret the various existing experimental
data on the single spin asymmetries in inclusive pion production
by the polarized proton and antiproton beams. As the basis of
analysis the chromo-magnetic string model is used. A whole
measured kinematic region is covered. The successes and fails of
such approach are outlined. The possible improvements of model are
discussed.

\end{abstract}

\section*{Introduction}

The growing amount of the experimental data on the single spin
asymmetry (SSA) attracts attentions of the high energy spin
community. This is understandable since the perturbative quantum
chromo-dynamics (pQCD) predicts a zero spin effect at the
asymptotic energy, while the experiments furnish up to now the
opposite signatures. For example, SSA shows to be nonzero at the
RHIC energy of $\sqrt{s}=200$ GeV \cite{starfpd}. Therefore the
different approaches, not so strongly justified as pQCD, but able
to explain all SSA effects are needed.\\
In 1988 the model was proposed \cite{rysk} for interpretation of
the SSA in inclusive pion productions and $\Lambda$ inclusive
polarization. The model was able to describe the scarce data
existing at that time. Those experimental data came up from the
unpolarized beams striking the polarized target and such
experiments suffer from the poor statistical accuracies. Since
later a lot of new data were published and mostly from the
interaction of the polarized beams with pure hydrogen target.
Recently the polarized Relativistic Heavy Ion Collider (pRHIC)
becomes operational and the first data on the SSA for inclusive
$\pi^0$ were published \cite{starfpd}. The PROZA experiment at
IHEP, Protvino, produced for the first time the data on
SSA in the polarized proton target fragmentation region
using unpolarized proton and pion beams. \\
The main goal of this article is to apply the chromo-magnetic
string model (CMSM) of the paper \cite{rysk} to the new
experimental data on the inclusive pion asymmetries. The CMSM
offers a simple analytical expression for SSA description. The
original paper was devoted mostly to the interpretation of the SSA
in inclusive $\pi^0$ production. In current paper we extend the
CMSM to the description of the SSA in the inclusive charged pion
production too.\\
In section 1 we remind the main features of CMSM according to
\cite{rysk}. The formulae of CMSM are specified for two distinct
kinematic regions: the central (CR) and beam fragmentation (BF)
region. Section 2 describes the application of the model to the
CR, while the section 3 is devoted to the BF region. Also in
Section 3 we expand the model to the SSA in the inclusive charged
pion productions. The unique data of E704 experiment on the SSA
obtained with the polarized antiproton beam are analyzed too in
scope of the isotopic spin and charge conservations. In summary we
outline the successes and fails of the CMSM in description of the
SSA in inclusive pion production at high energy. We make some
predictions and indicate the possible improvements of the CMSM model.\\
\section{\bf The chromo-magnetic string model}
The basis of this model is following. After collision of two
hadrons the color tube (string) is stretched between them. In
simple case this tube contains the flux of the chromo-electric
field. But such a system is not stable. In order to make it stable
the chromo-magnetic field should circulate around this tube. The
interaction of this field with the color magnetic moment of the
polarized quark leads to a kick transverse to the string axis.
This kick depends on the spin orientation of the polarized quark.
The estimate of this kick gives a magnitude of order $\delta
p_T\simeq 0.1$ GeV/c \cite{rysk}. If we introduce the invariant
differential cross section $\rho=E\frac{d\sigma}{d^3p}$ for
polarized quark asymmetry we get the relation
\begin{equation}\label{pa}
   A_{q}=\frac{\rho(+)-\rho(-)}{\rho(+)+\rho(-)},
\end{equation}
where (+) and (-) arguments in $\rho$  mean the up and down direction
of the quark polarization. Since $\delta p_T < p_T$, where $p_T$ is a
transverse momentum of the final pion to which the polarized quark
fragments we get the following left-right asymmetry in the quark
emission
\begin{eqnarray}\label{qas}
 A_q(x)=\frac{d\sigma(p_T +\delta p_T)-d\sigma(p_T-\delta p_T)}
 {d\sigma(p_T+\delta p_T)+d\sigma(p_T-\delta p_T)}= \nonumber \\
 \delta
 p_T\cdot \frac{\delta}{\delta p_T}(\frac{d\sigma}
{d^3p})/\frac{d\sigma}{d^3p}= {\delta p_T}\cdot B.
\end{eqnarray}
Here B is a slope parameter of $\rho$ as it is defined in standard
way.
\begin{equation}\label{bsl}
B=\frac{\delta}{\delta p_T}(\frac{d\sigma}
{d^3p})/\frac{d\sigma}{d^3p}=\frac{\delta}{\delta p_T}(ln{\rho}).
\end{equation}
The general formula for description of the SSA can be expressed in
the following form
\begin{equation}\label{an}
A_N(x) = P_q(x)\cdot A_q(x)\cdot w(x).
\end{equation}
Here $P_q(x)$ is a polarization of an initial quark, carrying a
portion x of the initial polarized proton momentum, $A_q(x)$ is so
called a quark "analyzing power" defined above (\ref{qas}); w(x)
presents a fractional contribution of a channel of interest in
parton-parton interactions.\\
The transverse quark polarization (transversity), $P_q(x)$, is
related to the spin dependent structure function $g_2(x)$ measured
in deep inelastic scattering (DIS). Unfortunately this function
has no parton interpretation. For the $100\%$ polarized on
mass-shell quark $g_2(x)=0$. On the other hand it looks reasonable
to expect that the x-behavior of $P_q(x)$ is similar to that of
$g_2(x)$. Note that the present data \cite{g2x} indicate the
threshold like form of the function $g_2(x)$ which becomes very
small for $x\leq 0.2$. Similar effect is seen in experimental data
for SSA in beam fragmentation region (see next chapters).
Therefore such effect should be included in a more precise
parametrization of $P_q(x)$ in future. Since at moment the
transversity can not be directly extracted from experimental data
one needs the theoretical inputs. According to the
non-relativistic quark model (NRQM) the transversity may be taken
in the following way
\begin{eqnarray}\label{nrqmp}
P_q(x)=\frac{2}{3}\cdot x \mbox{ for } \pi^+,
P_q(x)=\frac{1}{3}\cdot x \mbox{ for }\pi^0 \mbox{ and } \nonumber
\\ P_q(x)=-\frac{1}{3}\cdot x \mbox{ for } \pi^-.
\end{eqnarray}
So the NRQM predicts distinct effects for the pion SSA. First, the
sign of asymmetry is positive for $\pi^+ \mbox{ and for }\pi^0$,
while it is negative for $\pi^-$. Second, the magnitude of
asymmetry is expected to be higher for the $\pi^+$, than for
$\pi^0$ and $\pi^-$. These predictions are in general compatible
with experimental data, though
some deviations may happen.\\
Another theoretical inputs can be taken from paper \cite{ansel}
which aimed to explain the "spin crisis" by accounting the pion
cloud around the proton. Starting with NRQM, after account for
pion loops correction one gets $\Delta u$=0.79 (the sum for two
u-quarks), and $\Delta d$=-0.31 corresponding to $P_d=-0.8P_u$. In
such a case we can take $P_u(x)=0.7x$ and $P_d(x)=-0.55x$ leading
to
\begin{eqnarray}\label{ansr}
P_q(x)=0.7\cdot x \mbox{ for } \pi^+, P_q(x)=0.28\cdot x \mbox{
for } \pi^0 \mbox { and } \nonumber \\
P_q(x)=-0.55\cdot x \mbox{ for } \pi^-.
\end{eqnarray}
Comparing (\ref{ansr}) to (\ref{nrqmp}) one can see that the
difference is essential only for $\pi^-$ and this fact will be
taken into account at comparisons of the calculated SSA with
experimental data.\\
$A_N(x)$ may depend on $x_F$ which is
a Feinman parameter, on $p_T$ which is a transverse momentum of
pion, and on $s$ which is a square of the total energy in the
center of mass system of colliding particles.\\
The invariant differential cross section is assumed to be measured
before or simultaneously with SSA.\\
The general expression for weighting factor depends on the parton
distribution function $V^{a}(x)$ and parton fragmentation function
$D ^{a}(x)$ (a-parton flavor). Assuming that the polarization
information may be carried mostly through the polarized quark q
and much less by gluons we get the weighting (or asymmetry
diluting) factor through the following expression
\begin{equation}\label{wf}
    w(q)=\frac{\sigma(q)}{\sigma(q)+\sigma(g)}.
\end{equation}
Assuming that the main contribution comes from $qg'$ (for
$\sigma(q)$) and from $gg'$ (for $\sigma(g)$) scattering via the
t-channel gluon exchange between the gluon $g'$ from unpolarized
nucleon and quark $q$ or gluon $g$ from the polarized proton the
contributions from quark and gluon may be written as
\begin{eqnarray}\label{qgcs}
\sigma(q) \propto
C_F\int_x^1V^{q}(\frac{x}{z})D^{q}(z)\frac{dz}{z}, \nonumber \\
 \sigma(g) \propto C_A\int_x^1V^{g}(\frac{x}{z})D^{g}(z)\frac{dz}{z},
\end{eqnarray}
where the color factors $C_F=\frac{4}{3}$ and $C_A=3$, z is a fraction
of the polarized quark momentum carried by the final hadron.\\
The parton and gluon distribution functions were taken in the
forms
\begin{eqnarray}\label{pdf}
V^{q}(x)=x\cdot v(x)=2.8\cdot\sqrt{x}\cdot(1-x)^2,\nonumber \\
V^{g}(x)=x\cdot g(x)=3.0\cdot (1-x)^5.
\end{eqnarray}
The fragmentation functions for three types of inclusively
produced pions were taken in the forms:
\begin{eqnarray}\label{pff}
D^{u}_{\pi^+}(z)=\frac{4}{3}(1-z), D^{u}_{\pi^0}(z)=\frac{2}{3}(1-z),\
D^u_{\pi^-}(z)\approx 0; \nonumber \\
D^g_{\pi^+}(z)=D^g_{\pi^-}(z)=D^g_{\pi^0}(z)= (1-z)^2.\
\end{eqnarray}
The conservation of the isotopic spin and charge in parton
interaction leads to the relations
\begin{eqnarray}\label{icc}
D^u_{\pi^+}(z)=D^{\bar u}_{\pi^-}(z)= D^d_{\pi^-}(z)= D^{\bar
d}_{\pi^+}(z), \nonumber \\
D^u_{\pi^-}(z)=D^{\bar u}_{\pi^+}(z)= D^d_{\pi^+}(z)= D^{\bar
    d}_{\pi^-}(z).
\end{eqnarray}
The practical applications of these relations will be given below
in appropriate places.\\
The weighting factor (\ref{wf}) was calculated using the relations
(\ref{qgcs}), (\ref{pdf}), and (\ref{pff}). The final version of
function w(x) for different pions may be approximated by
\begin{equation}\label{wf+-0}
w(x)=\frac{\sqrt{x}}{\sqrt{x}+c(1-x)^{4.5}},
\end{equation}
where c is:$\ c_+=0.48$ for $\pi^+$, $c_0=0.64$ for $\pi^0$,
and $c_-=0.96$ for $\pi^-$.\\
 As it is seen from relations (\ref{an}) and (\ref{wf+-0}) the
analyzing power is calculated from $\rho \mbox { and } w(x)$
without any free parameter.\\
At lower energy, small $p_T$ and $x_F$ one may expect the
significant contributions to the pion yields from resonance
productions \cite{mt}. Therefore we arbitrary selected
experimental data with $\sqrt{s} \geq$ 10 GeV to avoid in some
extent this problem.
\begin{center}
\section{\bf Central region}
\end{center}
 At present we know 3 pieces of the published data on the SSA
for inclusive $\pi^0$ production in the central region (no data
for charged pions). They are:
\begin{enumerate}
\item Inclusive SSA for reaction
\begin{equation}\label{ppocr}
\underline{p(\uparrow) +p\rightarrow \pi^{0} +X}.
\end{equation}
 Experiment E704 at Fermilab measured this reaction at 200 GeV/c.
 Kinematic region:
$-0.15\leq x_F\leq 0.15$, $\  1.48\leq p_T(GeV/c)\leq 4.31$ \cite{E704cr}.\\
\item Inclusive SSA for reaction
\begin{equation}\label{appocr}
\underline{\bar p(\uparrow) +p\rightarrow \pi^{0} +X}.
\end{equation}
 Experiment E704 at Fermilab measured this reaction at 200 GeV/c.
 Kinematic region:
$-0.15\leq x_F\leq 0.15$,\ $ 1.48\leq p_T(GeV/c)\leq 3.35$ \cite{E704cr}.\\
\item Inclusive SSA for reaction
\begin{equation}\label{ppt70}
\underline{p +p(\uparrow)\rightarrow \pi^{0} +X}.
\end{equation}
Experiment PROZA-M at IHEP, Protvino, measured this reaction at 70
GeV/c. Kinematic region:
$-0.15\leq x_F\leq 0.15,\  1.05\leq p_T(GeV/c)\leq 2.74$ \cite{pp70cr}.\\
\end{enumerate}
In following we discuss these data in listed above order.\\
1. In E704 experiment up or down (to the horizontal plane)
polarized proton beam of 200 GeV/c striked the unpolarized proton
target. The inclusively produced $\pi^0$-s were detected by the
electromagnetic calorimeters around the central region, that is,
at the $\pi^0$ emission angle of $\approx 90^0$ in c.m.s. Since
the experimental invariant differential cross section, $\rho$, is
almost exponential in function of $p_T$ its slope parameter B is
practically constant. For example, at 200 GeV/c the slope
parameter B=(4.19$\pm 0.08) (GeV/c)^{-1}$ \cite{E704cr}. Then the
quark analyzing power, $A_q=0.1B=0.419$.\\
According to the previous discussions (\ref{nrqmp}) and
(\ref{ansr}) we take the transversity $P_q(x)=\frac{1}{3}x$,
constant factor $c_0$=0.64 (\ref{wf+-0}) and inserting all above
relations into (\ref{an}) we got the final result for $\pi^0$
asymmetry in the central region
\begin{equation}\label{anf}
A_N(x)=\frac{0.14\cdot {x}^{1.5}}{\sqrt{x}+0.64(1-x)^{4.5}}.
\end{equation}

\begin{figure}
\centerline{\epsfxsize=13cm\epsfbox{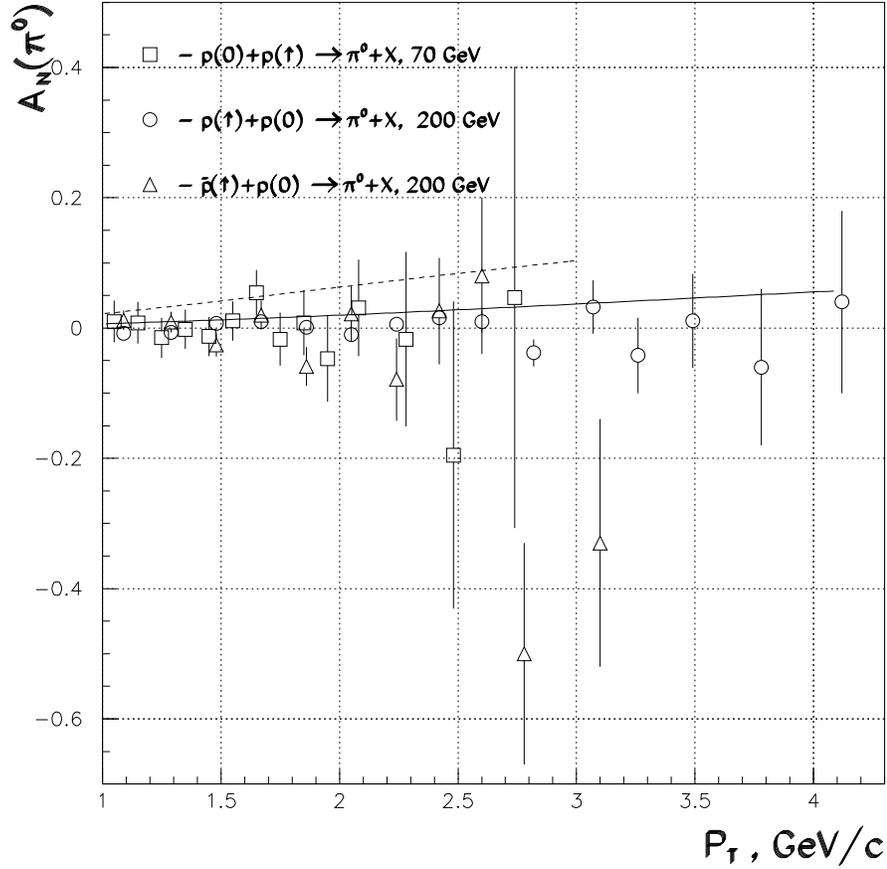}}
\caption{Asymmetry in function of $p_T$ in the central region for
reaction $p+p\rightarrow \pi^0+X$. At 70 GeV/c the target is
polarized, while at 200 GeV/c the beam is polarized. SSA is also
shown for reaction $\bar p(\uparrow)+p\rightarrow \pi^0+X$ at 200
GeV/c. The model predictions are ploted by the solid line for 200 GeV/c 
and by the dashed  line for 70 GeV/c.}
\end{figure}

One can see several consequences of this formula (\ref{anf}).
First, it is scaled on argument x= $x_T=\frac{2p_T}{\sqrt{s}}$
assuming that the slope parameter is constant. Second, it becomes
zero at x=0 mostly due to a quark polarization and in some extent
due to the gluon contribution (second term in denominator). Third,
when x approaches to one asymmetry increases as linear function of
x. All these behaviors remind the corresponding experimental
asymmetry.\\
The result of the model prediction according to the formula
(\ref{anf}) is presented in Fig.1 by the solid
line. This
prediction is consistent with the experimental data at 200 GeV/c.\\
2. The second group of data is relevant to the SSA in inclusive
$\pi^0$ production by the polarized antiproton beam of 200 GeV/c
(E704 experiment) (\ref{appocr}) \cite{E704cr}. This reaction
differs from (\ref{ppocr}) by charge of quarks, that is, all
quarks in polarized proton are replaced by the corresponding
antiquarks in polarized antiproton. Therefore the charge
conservation requires the equality of the SSA's in these two
reactions, namely
\begin{equation}\label{pi0}
A_N(\bar
p(\uparrow)+p\rightarrow\pi^0+X)=A_N(p(\uparrow)+p\rightarrow\pi^0+X).
\end{equation}
As seen (Fig. 1) this conclusion is consistent with the
experimental data in frame of error bars.\\
3. Third piece of new data \cite{pp70cr} is relevant to the
reaction (\ref{ppt70}). The data were taken at 70 GeV/c
unpolarized proton beam with polarized proton target. They are
presented in Fig.1 with opposite sign of asymmetry (to make a
comparison to the polarized beam case). The expression for
asymmetry at 70 GeV/c deduced from the CMSM looks like
\begin{equation}\label{cr70}
A_N(x_T)=0.2\cdot \frac{x_T^{1.5}}{\sqrt{x_T}+0.64(1-x_T)^{4.5}}.
\end{equation}
In comparison to (\ref{anf}) there are two changes. First is
relevant to the energy difference and second- to the different
slope parameters. At 70 GeV/c this parameter B=$(5.89\pm0.08)GeV^{-1}$
according to \cite{pp70cr}. The result of calculation is presented
in Fig.1 by the dashed 
line. In this case  one can see a
good consistency of prediction and experimental data. In order to
check the quantitative predictions of the CMSM one needs much better statistics.\\
4. $p(\uparrow) +p\rightarrow \pi^0 + X$, $\sqrt{s}=200$ GeV
(PHENIX experiment at RHIC). At moment there are no data on the
analyzing power of this reaction at $\sqrt{s}=200$ GeV. But PHENIX
collaboration recently published the precise data on the invariant
differential cross section of this reaction for $1\leq p_T\leq13$
GeV/c region at pseudorapidity $\mid \eta \mid\leq 0.39$
\cite{phenix}. This cross section was parameterized by authors in
the following form
\begin{equation}\label{dcs1}
\rho=E\frac{d^3\sigma}{dp^3}=A\cdot (1+\frac{p_T}{p_0})^{-n}.
\end{equation}
Here $A=386\  mb\cdot GeV^{-2}$, $p_0$=1.219 GeV/c and n=9.99. Now
we can calculate the slope parameter for the PHENIX invariant
differential cross section
\begin{equation}\label{slope1}
B(p_T)=\frac{dln{\rho}}{dp_T}=\frac{n}{p_0+p_T}.
\end{equation}
It is seen from (\ref{slope1}) that B($p_T$) decreases with growth of
$p_T$ as $p_T^{-1}$ and leads to the decrease of asymmetry at
large transverse momentum. This trend is similar to one stemming
out at the lower initial momentum ($\approx$ 200 GeV/c)
\cite{don}. The $\rho$ at momenta 100 and 200 GeV/c was presented
in the following form
\begin{equation}\label{dcs2}
\rho=E\frac{d^3\sigma}{d^3p}\propto (p_T^2 + M^2)^N
\cdot(1-x_T)^F.
\end{equation}
The fit to the experimental data leads to the values: $ N=-5.4\pm
0.2,\ M^2 = (2.3\pm 0.3)\  GeV^2 \mbox{ and }F=7.1\pm 0.4$, $x_T=
\frac{2p_T}{\sqrt{s}}$. Therefore one can find the slope parameter
\begin{equation}\label{slope2}
b(p_T)=\frac{2Np_T}{p_T^2+M^2}-\frac{2F}{\sqrt{s}(1-x_T)}.
\end{equation}
 The slope parameter at PHENIX (\ref{slope1}) decreases in the
 same way in function of $p_T$ as the slope parameter at around E704
 energy (\ref{slope2}) indicating on the independence of the slope
 parameter from the initial energy at $\sqrt{s}$=20-200 GeV energy
 region.\\
One can make the prediction for asymmetry in the above reaction
for PHENIX. It seems very small, less than 1\%. We hope to see
soon the experimental data from RHIC on $A_N(\pi^0)$.\\
Concluding the discussions of the SSA in central region we note
that better precisions in $A_N$ of order $\leq 1\%$ are needed for
testing the CMSM.
\begin{center}
\section{\bf Beam fragmentation region}
\end{center}
In this domain at energies $\sqrt{s}\ge$ 10 GeV there are the
following pieces of data:
\begin{enumerate}
\item The STAR result on the SSA in reaction
$p(\uparrow)+p\rightarrow\pi{^0}+X$  at $\sqrt{s}$=200 GeV/c.
Kinematic domain of measurement: $0.18\leq  x_F\leq 0.59,\ 1.5\leq
p_T(GeV/c) \leq 2.3,\  \mid \eta \mid \approx 3.8$ \cite{starfpd}.
\item  The E704 result on the SSA in reaction
$p(\uparrow)+p\rightarrow\pi{^0}+X$ at 200 GeV/c. Kinematic
domain: $0.03\leq x_F\leq 0.9,\ 0.5\leq p_T(GeV/c)\leq 2.0$
\cite{E704pp}. \item The E704 result on the SSA in reaction $\bar
p(\uparrow)+p\rightarrow \pi{^0}+X$ at 200 GeV/c. Kinematic
domain: $0.03\leq x_F\leq 0.67,\ 0.5\leq p_T(GeV/c)\leq 2.0$
\cite{E704bfr}. \item The E704 results on the SSA in reaction
$p(\uparrow)+p\rightarrow \pi^{\pm}+X$ at 200 GeV/c . Kinematic
domain of measurements:  $0.2\leq x_F \leq 0.9,\ 0.2\leq
p_T(GeV/c) \leq 1.5$\cite{E704PI+-}. \item The E704 results on the
SSA in reaction $\bar p\uparrow+p\rightarrow\pi^{\pm}+X$ at 200
GeV/c. Kinematic domain of measurements:  $0.2\leq x_F \leq 0.9,\
0.2\leq p_T(GeV/c) \leq 1.5$ \cite{bravar}.
\end{enumerate}
All listed above experimental data are shown in Fig.2 and Fig.3.\\
 The expression for asymmetry (\ref{an}) is still correct, as well as
for the weighting factor w(x). The main change consists in
replacing the argument x by $x_F$, which presents now the portion
of momentum acquired by hadron in the fragmentation of polarized
quark. SSA may be written in the following way (see (\ref{an}))
\begin{equation}\label{bfran}
A_N(x_F,p_T,s) = P_q(x_F)\cdot A_q(p_T, x_F, s)\cdot w(x_F,c).
\end{equation}
The factor c was defined earlier (see(\ref{wf+-0})).\\
In order to calculate $A_{q}$ we must know the slope parameter B
defined earlier (\ref{bsl}) for each reaction at the appropriate
kinematic domain. In following we discuss point by point
the above experimental data on SSA's in the beam fragmentation region.\\
1. STAR made a measurement of SSA and $\rho \ $ (invariant
differential cross-section) at $\sqrt{s}$=200 GeV and the
pseudorapidity $\mid \bar \eta \mid=3.8$. In order to extract the
slope parameter the $\rho $ measured at the same experiment as
function of $x_F$ was presented as the exponential function of
$p_T$ and a fit to the data resulted in the slope parameter
B=$(5.94\pm 0.016)\  GeV^{-1}$ with $\chi^2=27$ at number of
points =5. Sure the fit may be improved by adding a new parameter
quadratic in $p_T$ but for our goal the linear approximation in
$p_T$ is acceptable.\\
\begin{figure}
\centerline{\epsfxsize=13cm\epsfbox{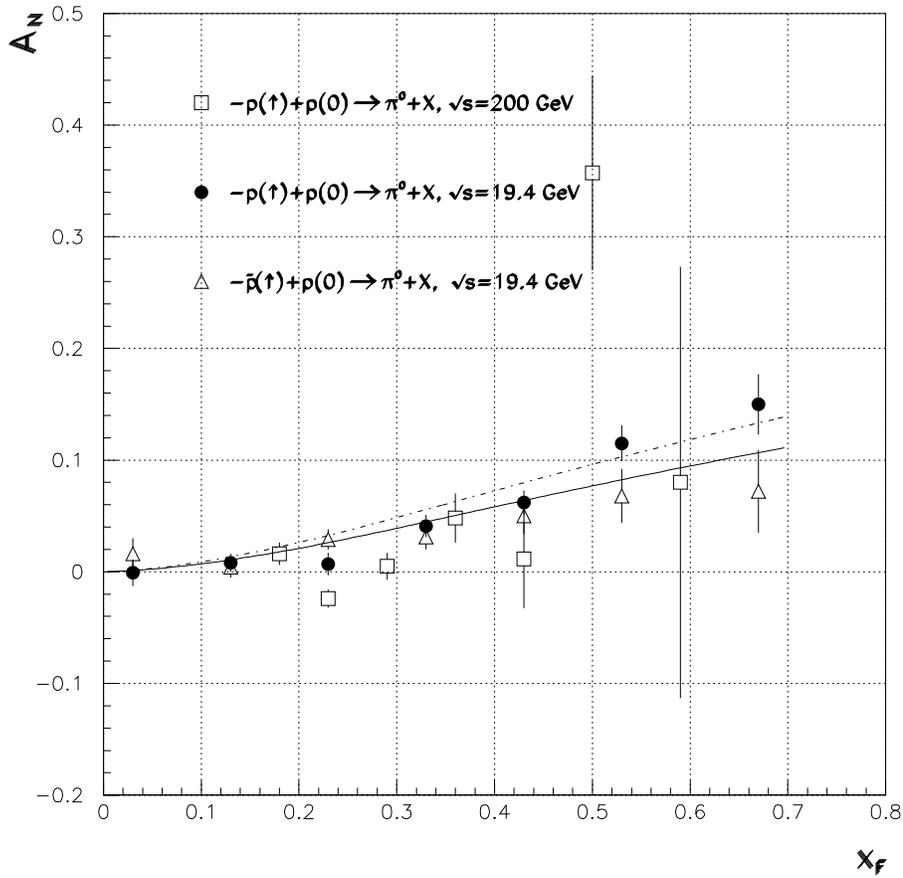}} 
\caption{Asymmetry in function of $x_F$ in the polarized beam
fragmentation region for reaction $ p (\uparrow) +p\rightarrow
\pi^0+X$ at $\protect\sqrt{s}$=19.4 and 200 GeV. The data for
reaction $\bar p(\uparrow) +p\rightarrow \pi^0+X$ are also shown.
The solid line is the CMSM prediction for $\protect\sqrt{s}$=19.4
GeV data, the dotted line is for $\protect\sqrt{s}$=200~GeV data.}
\end{figure}

So the final expression for $\pi^0\ $SSA at $\sqrt{s}$=200 GeV looks like
\begin{equation}\label{fpd}
A_N(x_F)=\frac{0.2\cdot x_F^{1.5}}{\sqrt{x_F}+0.64(1-x_F)^{4.5}}.
\end{equation}
The result of calculation is shown in Fig.2 by the dotted line and
it is consistent with
the first STAR experimental data.\\
2. Since the E704 collaboration did not make the direct measurement
 of the invariant differential cross section simultaneously with
 SSA measurement at 200 GeV/c we took such data from paper
 \cite{carey}. In this paper the $\rho$ was measured in the interval of
50-400 GeV/c initial proton momentum and for the different
laboratory angle of the photon emission $30\leq \theta_{L}(mrad)
\leq 275$. The measured photon spectrum was converted to
$\rho(\pi^0)$ by applying the Sternheimer's method. The transverse
momenta of $\pi^0$ were in interval 0.3-4.0 GeV/c. The closest to
our kinematic domain are data at laboratory $\pi^0$ production
angle $\theta_{L}$=30 mrad. We took data in the region
$0.6<p_T(GeV/c)<2$. By using the exponential function with slope
parameter B we found by fit that $B=(4.75\pm 0.07)\ (GeV/c)^{-1}$.
For 9 experimental points $\chi^2$/d.o.f.=0.7. Using the relation
$x_F=2p_T/(\sqrt{s}\cdot tan{\theta_{cms}})$ the $x_F$ region for
experiment was defined as 0.03-0.27. In frame of these data it is
impossible to go
farther in $x_F$. \\
Taking the quark polarization as $P_q(x_F)=(1/3)\cdot x_F$ the
final formula for SSA at 200 GeV/c (E704 experiment) for $\pi^0$
is
\begin{equation}\label{pi0e704}
 A_N(x_F)=\frac{0.16x_F^{1.5}}{x_F^{0.5}+0.64(1-x_F)^{4.5}}.
\end{equation}
The result of calculation is presented in Fig.2 as a solid line.
One can see a qualitative consistency between the model prediction
and the data of E704 experiment.\\
3. In the same Fig.2 we present the SSA data for
reaction $\bar p(\uparrow)+p\rightarrow \pi{^0}+X$ at 200 GeV/c.
The model prediction is simplified by using the relation (\ref{pi0})
which is a consequence of transformation from quarks to antiquarks.
As seen from Fig.2 there is a good consistency of model prediction
(solid line) with experimental data.\\
 4. Figure 3 presents the experimental data from E704 experiment on the SSA for
the inclusively produced pions at 200 GeV/c in the following reactions:
\begin{equation}
p(\uparrow)+p\rightarrow \pi^{\pm,0}+X,\ \ \\
 \bar p(\uparrow)+p\rightarrow\pi^{\pm,0}+X.
\end{equation}

For calculation of the quark analyzing power, $A_q$, we must estimate
the slope parameters B$^+$ for positive and B$^-$ for negative pion
productions. This task was done by using the data from paper
\cite{break}. The invariant differential cross-section was taken
at $\sqrt{s}=31$ GeV, closest to E704 data. The slope parameters
were estimated by taking two values of $ p_T$ =0.82 GeV/c and $
p_T$ =1.42 GeV/c and averaged over region $0.35<x_F<0.65$. Finally
we took $B^+$=5.55 GeV$^{-1}$ and $B^{-}$=5.33 GeV$^{-1}$. So the
expression for $\pi^+$ SSA taking $P_q(x)=(2/3)\cdot x$ looks like
\begin{equation}
A_N^{\pi^+}(x_F)=
\frac{0.37x^{1.5}_F}{x_F^{0.5}+0.48(1-x_F)^{4.5}}.
\end{equation}
Prediction of this formula for SSA in inclusive $\pi^+$ production
by polarized protons is presented in Fig. 3a by the solid line.
Data (closed circles) confirm this prediction fairly well excluding
the region $x_F<$0.4. \\
Fig.3b shows the SSA for inclusive $\pi^-$ production in pp collisions
at 200 GeV/c. According to the CMSM taking $P_q(x)=-(1/3)\cdot x$
the following relation for $\pi^-$ SSA can be written
\begin{equation}
A_N^{\pi^-}(x_F)=-
\frac{0.18x^{1.5}_F}{x_F^{0.5}+0.96(1-x_F)^{4.5}}.
\end{equation}
Following the non-relativistic quark model we assumed the negative
sign of the d-quark polarization. As seen from Fig.3b the CMSM
prediction (solid line) follows qualitatively the experimental
data (closed circles) but it does not give a quantitative
agreement. A better description of SSA at the large $x_F$
furnishes the dotted line corresponding to the relation
$P_q(x)=-0.55x$ stemming out from paper \cite{ansel}. But in both
cases the model needs some improvements in order to describe
better the smallest $x_F$
region. \\
 5. Fig.3c presents the SSA versus $x_F$ for inclusive $\pi^0$
productions by the polarized proton (closed circles) and
antiproton (open circles) beams. As seen from this Fig.3c the
charge conjugation rule (\ref{pi0}) is well respected by the
experiment. For SSA in the inclusive charged pion productions the
following charge conjugation rule  should be respected
\begin{equation}\label{anbpp}
A_N(\bar p(\uparrow)+p\rightarrow \pi^{\pm}+X)\approx
A_N(p(\uparrow)+p\rightarrow \pi^{\mp}+X).
\end{equation}

\begin{figure}
\centerline{\epsfxsize=14cm\epsfbox{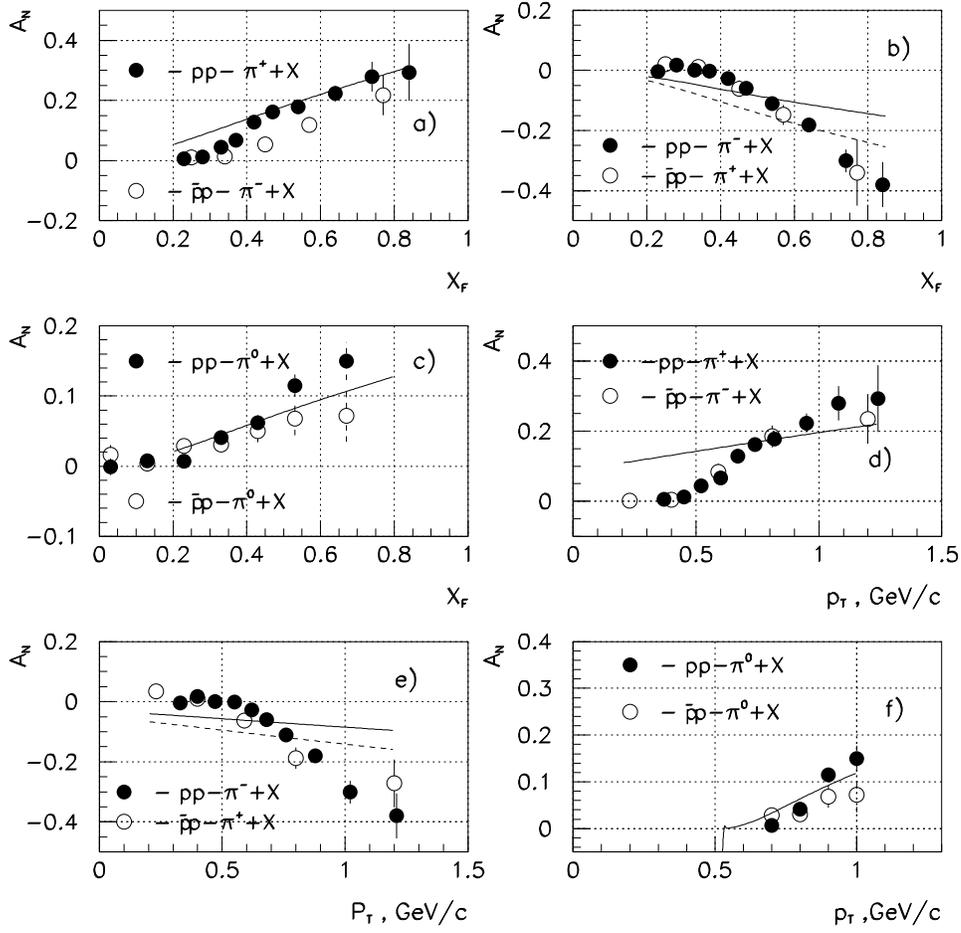}}
 \caption{
Test of the charge conjugation rules (\protect\ref{pi0}) and
(\protect\ref{anbpp}) for SSA in inclusive pion productions in
$p(\uparrow)p-$ (closed circles) and $\bar p(\uparrow)p$-(open
circles) collisions at 200 GeV/c. SSA are given in function of
$x_F$ in Figures a), b), c) and in function of $p_T$ in Figures
d), e), and f). Solid lines are the CMSM predictions for the
initial quark polarizations (\protect\ref{nrqmp}) and the dashed
lines are for conditions (\protect\ref{ansr}).}
 \end{figure}

According to this relation one can predict the inclusive charged
pion asymmetries produced by the polarized antiproton beam without
any free parameter. The comparison of model prediction with
experimental data is illustrated in Fig.3. The  solid lines are
model predictions. The experimental data for pp-collisions are
shown by the closed circles, while data for $\bar pp$ collision
are indicated by the open circles. Relation (\ref{anbpp}) is
consistent with data in general (Fig.3) but there are the drastic
discrepancies in some cases  which are not yet understood. For
example, Fig.3a shows, that the experimental data violate the rule
(\ref{anbpp}), while model gives a fairly well description of the
SSA for $\pi^+$ production. In Fig.3b the model (\ref{ansr}) gives
a better description of data than parametrization (\ref{nrqmp}).
The data in Fig.3c are consistent with expectation and the CMSM
gives a good description of SSA.

 The Fig.3d, 3e, and 3f show the $p_T$ dependencies of the SSA for
$\pi^{\pm,0}$ inclusively  produced by protons and antiprotons. In
the explicit form such dependencies were presented in the original
papers only for reactions $\bar p(\uparrow)p\rightarrow
\pi^{\pm}X$ \cite{bravar} (Fig.3d and Fig.3e, open circles). Other
figures were restored by us using the tables of data in the
original papers \cite{E704pp}, \cite{E704bfr}, and
\cite{E704PI+-}. In function of $p_T$ all SSA are consistent with
the charge conjugation rule (compare the closed and open circles
in Fig.3d, 3e, and 3f).

The solid lines are predictions of the CMSM.  These lines were
calculated by using the relation $x_F\approx a+bp_T$ deduced from
the tables of data in the original papers. Constants a and b were
defined by fit for each pair of SSA as they are presented in
figures. The consistencies between data and predictions are
reasonable, though in Fig.3d and Fig.3e the discrepancies are seen
at low $p_T$ between model prediction and the experimental data.
At such low $p_T$ some non-perturbative interaction ( in
particular, instanton induced)
 may affect  our predictions by changing the weighting factor w(x)
 and/or by flipping quark helicity.

Concluding this section we note that in general the CMSM gave the
qualitative description of most of the SSA effects in a whole
measured kinematic domain. Nevertheless it fails in the
interpretations of the one feature of SSA, namely, the evidence of
$A_N\approx 0$ in the interval $0<x_F<0.4$. This is illustrated by
the $A_N(p(\uparrow)p\rightarrow\pi^-+X)$ (see Fig.3b) and
$A_N(\bar p(\uparrow)p\rightarrow\pi^++X)$ (see Fig.3d). This
event might be related to some threshold effect like resonance
production.
\section*{\bf Conclusion}
We propose a Chromo-Magnetic String Model producing the simple
analytic expressions for the single spin asymmetries of inclusive
charged and neutral pion productions. It describes all known
experimental data in the central and beam fragmentation regions.
The following experimental facts are naturally explained on the
basis of the parton model, conservation of the charge and the
invariance of the parton interaction with respect to the charge
conjugation: $A_N(\bar p(\uparrow) +p \rightarrow
\pi^{\pm}+X)\approx A_N(p(\uparrow)+p\rightarrow\pi^{\mp}+X)$ and
$A_N(\bar p(\uparrow)+p\rightarrow \pi^{0}+X)\approx
A_N(p(\uparrow)+p\rightarrow \pi^{0}+X)$. The qualitative
descriptions of the $p_T$, $x_F$ and energy dependencies of the
single spin asymmetry seem reasonable. The CMSM should be applied
beyond the domain of resonance production and out of the threshold phenomena.\\
To improve the model one has to use the quark and gluon
distributions ($V(x)$) given by the global parton analysis and the
fragmentation functions ($D(z)$) measured in $e^-e^+$-annihilation
together with the NLO expressions for the hard parton-parton
scattering cross sections. This will provide a more precise
weighting factor $ w(x)$ (\ref{wf}). \\
Next (and more important) is to use a more realistic formula for
the quark polarization $P_q(x)$. Unfortunately we can not use the
spin dependent structure function $g_2(x)$ measured in the deep
inelastic scattering (DIS) directly. This function has no parton
interpretation and for 100$\%$ polarized on mass-shell quark
$g_2(x)$=0. Therefore, the value of $g_2(x)$ measured in DIS  is
expected to be smaller than $P_q(x)$. On the other hand it looks
reasonable to choose $P_q(x)=c\cdot g_2(x)$ with such a
coefficient  'c' that  (according to "spin crisis") the whole
polarization carried by valence quark will be about 1/4 of the
incoming proton polarization. In this way, thanks to the $x$
behavior of $g_2(x)$, we hope to reproduce the behavior of SSA
observed in the data at $x_F\leq 0.3$.\\
 We expect that the coming soon data from the polarized RHIC will test
our model more precisely.\\

We would like to express our thanks to L. Bland, Yu.V. Kharlov,
V.V. Mochalov and D.V. Sidorov for help and useful discussions.

\end{document}